\documentclass[journal]{IEEEtran}
\IEEEoverridecommandlockouts

\usepackage{cite}
\usepackage{amsmath,amssymb,amsfonts}
\usepackage{algorithmicx}
\usepackage{graphicx}
\usepackage{svg}
\svgsetup{
  inkscape=false,   
  inkscapepath=svgdir  
}
\usepackage{textcomp}
\usepackage{siunitx}
\usepackage{booktabs}
\usepackage{xcolor}
\usepackage{booktabs}
\usepackage{siunitx}
\usepackage{algpseudocode}
\usepackage[linesnumbered,ruled,lined]{algorithm2e}
\usepackage{placeins} 
\usepackage[utf8]{inputenc}     
\usepackage[T1]{fontenc}
\usepackage{subcaption}        
\usepackage{multirow}
\usepackage{makecell}          
\usepackage{microtype}         
\usepackage{enumitem} 
\usepackage{tabularx}
\usepackage{array}
\usepackage{float}
\usepackage{cuted}
\usepackage{url}
\usepackage{pgffor}  
\usepackage{graphicx}
\usepackage{stfloats}
\usepackage{needspace} 
\usepackage{url}
\usepackage{hyperref}
\newcommand{\sep}{1em}
\DeclareSIUnit{\sample}{sample}

\def\cwpowers{-10,-20,-30,-40,-50}   
\def\qppowers{-10,-20,-30,-40,-50}   
\def\imgwd{0.22\textwidth}              

\def\BibTeX{{\rm B\kern-.05em{\sc i\kern-.025em b}\kern-.08em
    T\kern-.1667em\lower.7ex\hbox{E}\kern-.125emX}}
\begin{document}

\title{Co-Channel Interference Mitigation Using Deep Learning for Drone-Based Large-Scale Antenna Measurements}

\author{%
  Kadyrzhan~Tortayev, Oliver Falkenberg Damborg, Jònas À Hàlvmørk Joensen, Jonas Pedesk, Yifa Li, Fengchun Zhang,  Zeliang An, Yubo Wang and Ming Shen%
  \thanks{All authors are with the Department of Electronic Systems, Aalborg University, Aalborg, Denmark. (e-mail: {ktorta24, odambo20, jjoe21, jpedes20}@student.aau.dk and {yifal, fz, yubow, mish}@es.aau.dk, and anzeliang1993@gmail.com
)
  
  Corresponding author: Ming Shen.}%
}

\maketitle

\begin{abstract}

Unmanned aerial vehicles (UAVs) enable efficient \textit{in-situ} radiation characterization of large-aperture antennas directly in their deployment environments. In such measurements, a continuous-wave (CW) probe tone is commonly transmitted to characterize the antenna response. However, active co-channel emissions from neighboring antennas often introduce severe in-band interference, where classical FFT-based estimators fail to accurately estimate the CW tone amplitude when the signal-to-interference ratios (SIR) falls below -10 dB. This paper proposes a lightweight deep convolutional neural network (DC-CNN) that estimates the amplitude of the CW tone. The model is trained and evaluated on real 5~GHz measurement bursts spanning an effective SIR range of --33.3~dB to +46.7~dB. Despite its compact size ($<$20k parameters), the proposed DC-CNN achieves a mean absolute error (MAE) of 7\% over the full range, with $<$1~dB error for SIR~$\geq$~--30~dB. This robustness and efficiency make DC-CNN suitable for deployment on embedded UAV platforms for interference-resilient antenna pattern characterization.
\end{abstract}


\section{Introduction}

 Accurate antenna‐pattern estimation is essential for the design, optimisation, and regulatory compliance of modern wireless systems \cite{ITU_S524}, \cite{SES_OPS}. Traditional antenna measurements are often conducted in anechoic chambers or compact test ranges, where absorptive walls suppress reflections to emulate free-space conditions. These facilities allow precise control of the measurement environment, enabling accurate far-field or near-field characterization without external interference. However, their physical dimensions must be several wavelengths larger than the antenna under test (AUT) to avoid edge diffraction, making them impractical or prohibitively expensive for very large apertures. Additionally, chambers cannot reproduce the AUT’s operational environment, meaning real-world effects such as multipath propagation, weather, and co-channel interference are absent from the measurements \cite{RoE2023Roadmap}, \cite{Kandregula2024UAV}.  Consequently, in‑situ measurements performed with the antenna operating in its native environment are increasingly required.

Uncrewed aerial vehicles (UAVs) enable such in-situ antenna characterization by providing precisely controlled trajectories, allowing for radiation pattern sampling of large antennas from predefined spatial positions \cite{UAVPropagation}.  Compared with mast or crane solutions, UAV surveys are faster, cheaper, and can be scheduled outside regular maintenance windows \cite{Kandregula2024UAV}.  In most UAV-based antenna measurement campaigns, the probe antenna is positioned significantly closer to the AUT than in conventional far-field test ranges. While far-field and compact-range facilities typically maintain distances that greatly exceed the radiating near-field boundary, UAV surveys often operate within or near this region to improve link budget and reduce the required transmit power. This close range operation enhances measurement SNR and enables the use of low-power CW probes, but also increases susceptibility to in-band interference from nearby active emitters making interference mitigation a critical requirement for reliable \textit{in-situ} pattern characterization. Because measurements are executed with the host site in full operation, neighboring antennas radiate co-channel emissions that couple into the probe link. In live LTE UAV measurements, signal-to-interference ratios have been observed as low as $-5.9\,$dB at 50 m altitude and $-7.8\,$dB at 100 m altitude \cite{Cai2019}. Under such conditions, conventional fast Fourier transform (FFT) techniques \cite{Mao2008NIC, Jacobsen2007ThreeBin, same2020}  such as three bin averaging or band-pass filtering  fail to provide reliable tone-gain estimates, resulting in distorted far-field patterns.

\begin{figure}[t]
    \centering
    \includesvg[width=\linewidth]{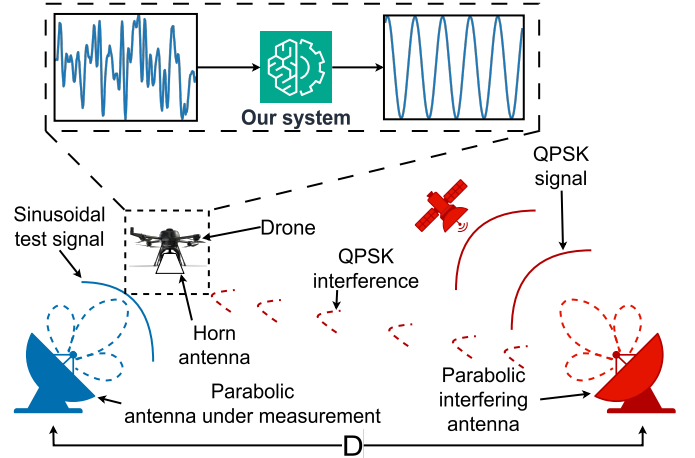}
    \caption{Overview of the measurement scenario and signal model.}
    \label{fig:FinalConceptFigure}
\end{figure}

Machine learning (ML) approaches have recently shown strong potential in mitigating such interference by operating directly on raw time-domain I/Q data \cite{Robinson2023ICC, Malolan2025Thesis, Zhao2021StackedLSTM, jayashankar2024, kim2023, Liu2024SIC}. These methods offer an opportunity to bypass classical frequency domain limitations and learn more robust representations. However, most prior ML works have focused on classification or denoising tasks, rather than quantitative estimation.
Robinson et al. \cite{Robinson2023ICC} train a CNN to detect and localise narrow-band jammers in Wi-Fi channels; Malolan’s thesis \cite{Malolan2025Thesis} employs a CNN-autoencoder to reconstruct QPSK payloads in the presence of single-tone and wide-band interferers; and Zhao et al. \cite{Zhao2021StackedLSTM} use a stacked-LSTM to separate aliased satellite signals, reporting SISNR gains up to 38 dB. All three works output a class label or a denoised waveform. In contrast, we tackle a regression task: estimating the amplitude gain of a known CW probe at SIR levels down to -40 dB and converting that gain into an antenna radiation pattern sample.

Our results indicate that lightweight, interference‑aware deep learning unlocks practical UAV‑based in‑situ antenna pattern measurement, bridging the gap between laboratory accuracy and field deployability.
Section II details the DC‑CNN architecture and training strategy.  Section III describes the UAV measurement campaign and the dataset.  Section IV compares DC‑CNN with FFT, and three heavier CNN/LSTM baselines.  Section V discusses the results, and Section VI concludes the paper with future work.



\begin{figure*}[!tb]
  \centering
  \makebox[\textwidth][c]{%
    \begin{subfigure}[b]{0.4\textwidth}
      \centering
      \includegraphics[width=\linewidth]{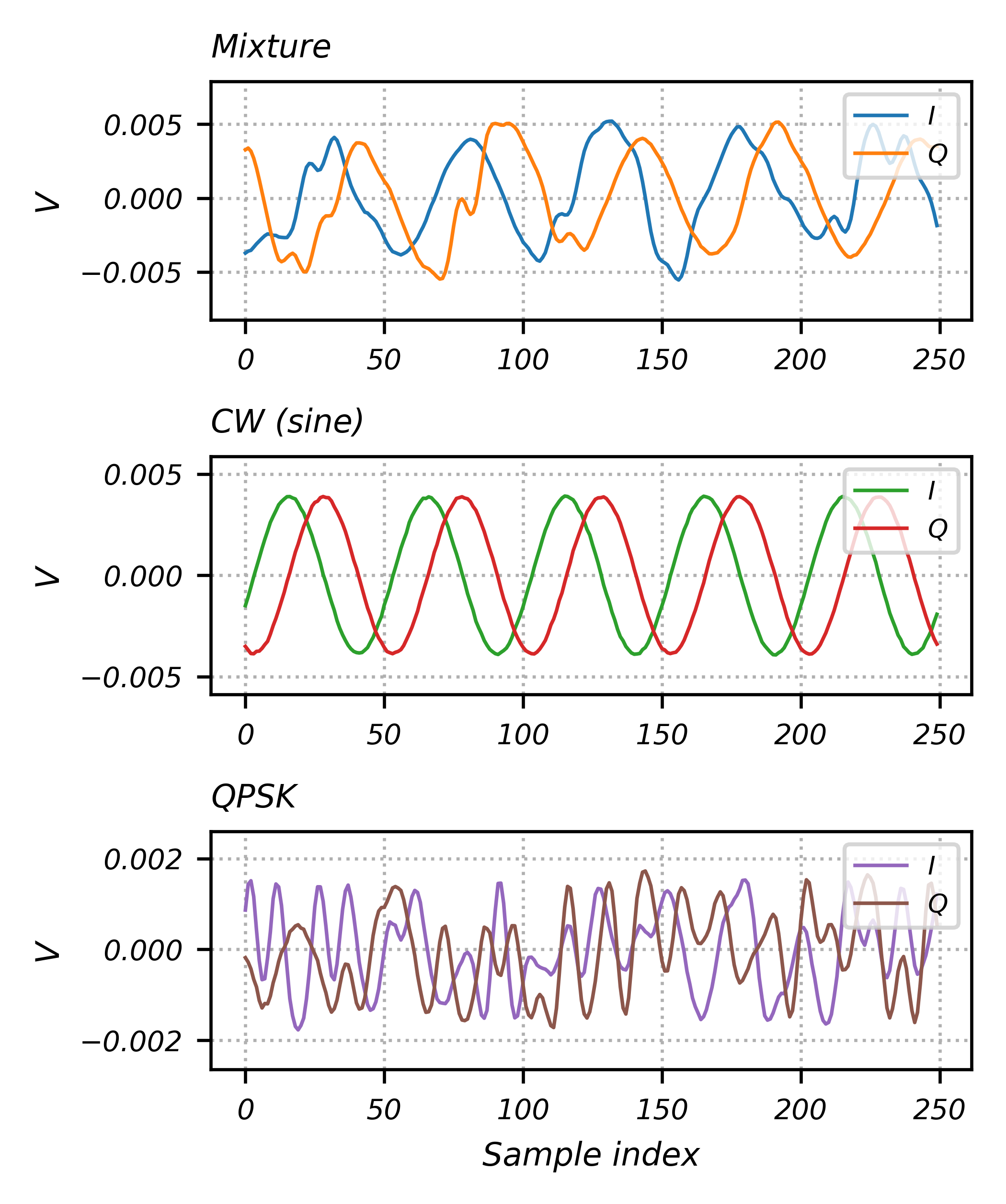}
      \caption{}
      \label{fig:inband_time}
    \end{subfigure}\hspace{0.05\textwidth}%
    \begin{subfigure}[b]{0.4\textwidth}
      \centering
      \includegraphics[width=\linewidth]{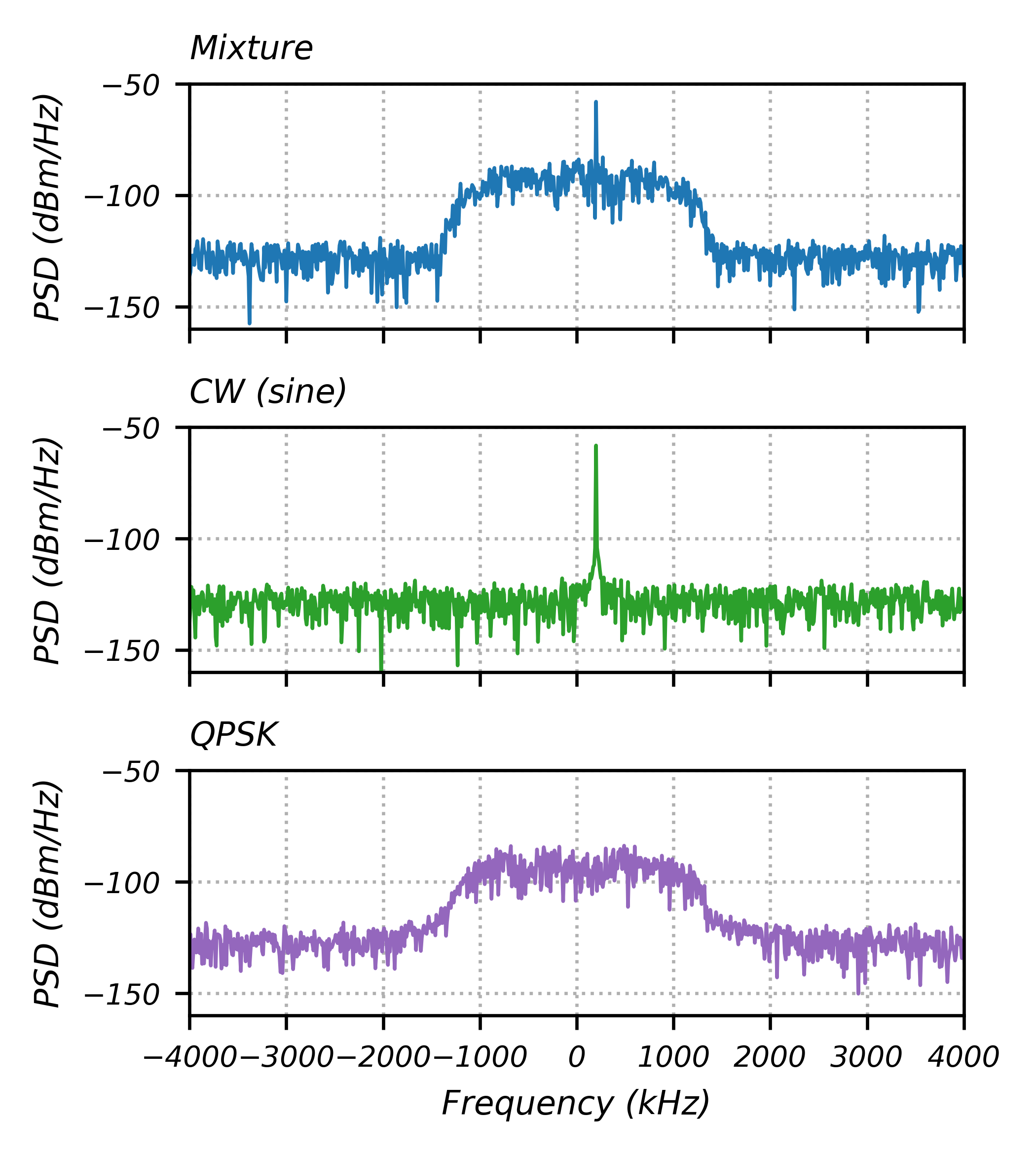}
      \caption{}
      \label{fig:inband_spec}
    \end{subfigure}%
  }
  \caption{Time-domain I/Q samples and their corresponding magnitude spectra. (a) Time-domain $I/Q$ samples of Mixture, CW(sine), and QPSK burst with the 200 kHz carrier offset shifted back. (b) Corresponding magnitude spectra (Hann window, 4096-point FFT).}
  \label{fig:inband_inspec}
\end{figure*}

\section{Architecture \& Method}\label{sec:method}

\subsection{Overview of DC‑CNN}
\begin{figure*}[!tb]
    \centering
    \includesvg[width=1.0\textwidth]{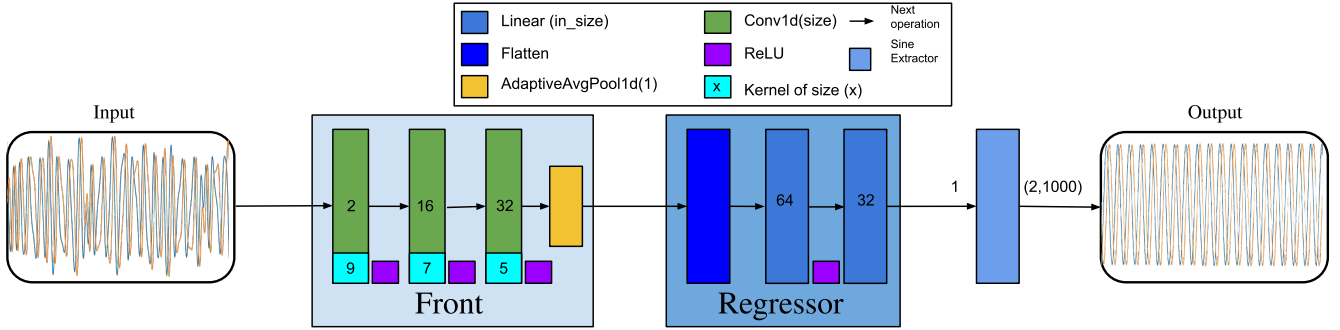}
\caption{{Three-layer CNN front-end (non-dilated).}\; 
Kernel sizes \{9, 7, 5\} yield a receptive-field \emph{radius} of 9 samples
(19 total), i.e.\ \SI{\pm0.9}{\micro\second} at the \SI{10}{\mega\sample\per\second} rate.
Adaptive-average pooling collapses the 1000-sample burst to a 64-dimensional
embedding; a two-layer MLP then regresses $\Re\{g\}$ and $\Im\{g\}$.
All layer dimensions are shown in blue.}
    \label{fig:BeaconPowerCNN}
\end{figure*}

As a baseline, we implement a compact DC-centered convolutional neural network (DC-CNN), where ``DC'' indicates that the CW probe tone is centered at $0$~Hz in the receiver baseband. As illustrated in Fig.~\ref{fig:inband_inspec} and Fig.~\ref{fig:BeaconPowerCNN}, the model receives a burst of complex in-phase and quadrature (I/Q) samples, reshaped into two separate channels, and processes them through three one-dimensional convolutional layers with progressively smaller kernel sizes ($9$, $7$, and $5$ samples), each followed by a rectified linear unit (ReLU) activation. Symmetric padding and a stride of $1$ preserve temporal alignment, allowing each layer to access both past and future context. The output of the final convolutional layer is reduced via adaptive average pooling to a 64-dimensional embedding, which is then passed to a two-layer fully connected regressor that outputs the real and imaginary components of the predicted CW amplitude.

To assess whether introducing a frequency offset improves interference robustness, we evaluate a Sine-CNN variant of the DC-CNN architecture. In this case, the CW probe tone is frequency-shifted by \SI{200}{\kilo\hertz} before training and inference, while all other architectural and training parameters remain identical to DC-CNN. This offset was chosen to ensure the probe remains within the recorded baseband while occupying a distinct spectral band from the QPSK interferer, as illustrated in Fig.~\ref{fig:inband_spec}. The comparison tests whether a nonzero probe frequency makes the target tone more distinguishable from in-band QPSK interference.

\subsection{Baseline Models}

To benchmark the performance of the proposed DC-CNN, we compare it with a set of baseline methods spanning deep learning and classical signal processing. Given the strong performance of long short-term memory (LSTM) networks in modeling sequential data, we evaluate multiple LSTM-based configurations alongside hybrid convolutional–recurrent models and a classical FFT estimator, as detailed below.

\textbf{DC-CRL} - A hybrid \emph{convolutional–recurrent} architecture designed to capture both short- and long-range temporal dependencies. It consists of two parallel convolutional encoders:  
(i) a \emph{wide-path} encoder with four stacked 1D residual blocks using exponentially increasing dilations $(1, 2, 4, 8)$ to enlarge the receptive field, and  
(ii) a \emph{narrow-path} encoder with two residual blocks at lower dilation rates to preserve fine-grained temporal detail.  
The outputs of the two paths are concatenated and passed through a unidirectional long short-term memory (LSTM) layer with 64 units, followed by a linear regression head that estimates the real and imaginary components of the CW amplitude.

\textbf{Bi-LSTM} - A recurrent architecture comprising two stacked bidirectional LSTM layers with 128 units per direction, designed to capture temporal dependencies in both forward and backward time. Between the LSTM layers, squeeze-and-excitation (SE) blocks reweight temporal feature channels to emphasize informative patterns. An initial 1D convolution projects the input I/Q samples into the hidden feature space, and the output sequence is collapsed via temporal averaging before passing to a linear regression head that estimates the real and imaginary components of the CW amplitude.


\textbf{CausalLSTM-SingleOutput} - A causal, unidirectional variant of the Bi-LSTM model. Causality is enforced by replacing the initial convolution with a causal convolution, ensuring predictions depend only on past and present samples. The model produces a single CW amplitude estimate for each input burst, making it suitable for low-latency streaming without look-ahead.

\textbf{CausalLSTM} - A sequence-to-sequence variant of CausalLSTM-SingleOutput that predicts the full 1000-sample CW waveform rather than a scalar amplitude. It is trained to match the clean sinusoidal target, allowing the network to perform fine-grained reconstruction under in-band interference. This task differs from scalar regression and is intended to assess the model’s ability to recover the complete time-domain signal.

\textbf{FFT 3-bin} - A classical baseline that applies a 4096-point FFT to the input burst and averages the squared magnitudes of the three frequency bins centered on $f_{\mathrm{CW}}$. Given the sampling rate in our setup, each FFT bin corresponds to approximately 7\,kHz, such that the three-bin window covers roughly 21\,kHz around the CW tone. For $\mathrm{SIR} < -12$\,dB, accuracy degrades sharply due to in-band interference energy leaking into adjacent bins.

\begin{table}[t]
  \caption{Trainable parameter counts for all evaluated models.}
  \label{tab:modelcomplexity}
  \centering
  \begin{tabular}{lr}
    \toprule
    \textbf{Model} & \textbf{Parameter Count} \\
    \midrule
    Bi-LSTM & 335,768 \\
    CausalLSTM & 385,648 \\
    DC-CRL & 154,786 \\
    DC-CNN & 16,370 \\
    \bottomrule
  \end{tabular}
\end{table}

All models were trained on the same dataset and hyper-parameters. They are  publicly available in our GitHub repository \cite{GitRepository}. To investigate the relationship between model size and estimation accuracy, we recorded the number of trainable parameters for each tested architecture. As shown in Table~\ref{tab:modelcomplexity}, model capacities range from under 20\,k parameters for lightweight CNNs to over 300\,k for recurrent architectures. This variation in model capacity enables us to assess whether increasing the number of trainable parameters improves regression accuracy in CW amplitude estimation.

\subsection{Training Procedure}

All models are trained on 8{,}000 real signal bursts collected using a Rohde\&Schwarz FSQ-26 spectrum analyzer, with 15\% of the data held out for validation. Each burst contains a CW probe mixed in hardware with a QPSK-modulated interferer. No synthetic mixing or on-the-fly augmentation (e.g., additive noise or interference) is applied; the I/Q samples are taken directly from physical measurements. In addition, 8{,}000 clean CW bursts (recorded without interference) serve as ground-truth references for training and evaluation.

The complex CW amplitude $g$ (defined in Section~\ref{sec:dataset}) is represented as a two-dimensional real vector $(\Re\{g\}, \Im\{g\})$, corresponding to the in-phase and quadrature components. Training is conducted for 200~epochs using the AdamW optimizer with a fixed learning rate of $2 \times 10^{-4}$ and a batch size of 16. Although the models estimate the full complex amplitude, supervision is applied only to the CW power, with the phase left unsupervised. This reflects our primary focus on amplitude-only \emph{pattern characterization}, though the predicted phase remains stable across the entire SIR range. 

The training objective minimizes the mean squared error between predicted and ground-truth CW power in the logarithmic (dB) domain:
\begin{equation}
    \mathcal{L}_{\text{MSE}} = \frac{1}{B} \sum_{i=1}^B \left( P(\hat{g}_i) - P(g_i) \right)^2 \quad [\text{dB}^2]
\end{equation}
where $B$ is the batch size, $\hat{g}_i$ is the predicted complex amplitude, $g_i$ is the ground truth, and $P(\cdot)$ denotes the conversion from complex amplitude to power in dB.



\section{Measurement \& Dataset Preparation}
\label{sec:dataset}

\subsection{Capture Setup}

\begin{figure}[H]
  \centering
  \begin{subfigure}[b]{\linewidth}\centering
 
    \includegraphics[width=.98\linewidth]{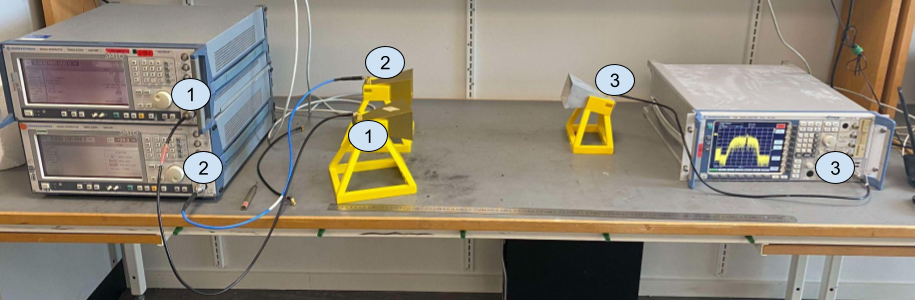}
    \caption*{\small(a) Photograph of the measurement bench (300 dpi)}
  \end{subfigure}\\[4pt]
  \begin{subfigure}[b]{\linewidth}\centering
    \includegraphics[width=.98\linewidth]{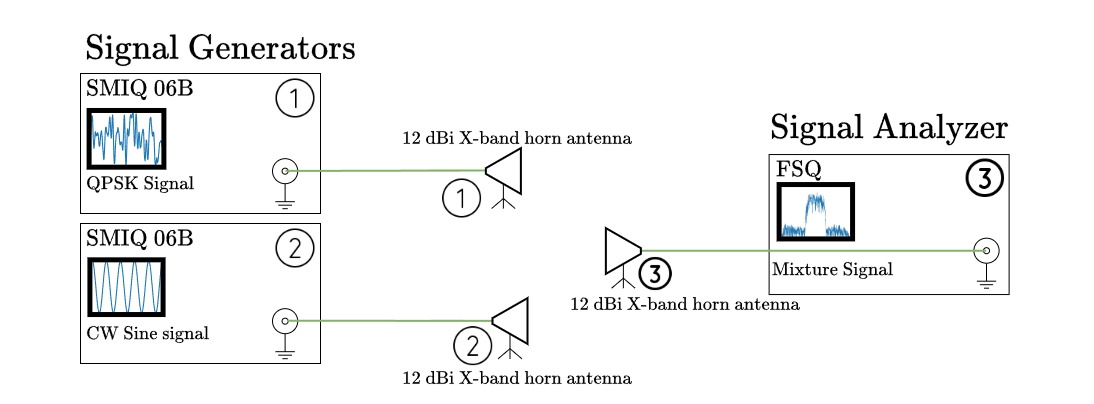}
  \end{subfigure}
   \caption*{\small(b) RF-signal-flow schematic}
  \caption{Laboratory capture bench used to emulate UAV near-field
           measurements.  SG: signal generator; SA: signal analyser.}
  \label{fig:setup}
\end{figure}

Two asynchronous R\&S SMIQ06B vector signal generators (SG) feed
a pair of \SI{12}{dBi} X-band horn antennas that illuminate a third,
identical horn positioned \SI{0.6}{\metre} in front of the array under
test (Fig.~\ref{fig:setup}).  The receive horn connects to a
R\&S FSQ-26 signal analyser (SA), which records complex baseband
at \SI{10}{MS/s}.  While the SGs and SA share an external \(10\;\text{MHz}\)
reference, they are \emph{not time-synchronised}; absolute phase is
therefore lost but this is acceptable because our regression target is
CW power, not phase. Given the 0.6\,m separation is within the radiating near-field ($R_{\mathrm{ff}} \approx 0.8\,\mathrm{m}$), the received field is strictly spherical. However, for the short bursts and narrow aperture considered here, we approximate the wavefront as planar in our models to simplify processing.

\subsection{Dataset and RF Configuration}

Table~\ref{tab:rfparams} summarizes the key RF parameters of the campaign. The system operates at a center frequency of \SI{5.000}{GHz}, with a CW tone offset by \SI{200}{kHz} from a \SI{4}{MHz}-wide QPSK interferer. Eight CW transmit power levels 
($P_\mathrm{CW} \in \{-10, -20, -25, -30, -35, -40, -45, -50\}\,\mathrm{dBm}$) are paired with five QPSK transmit power levels 
($P_\mathrm{QPSK} \in \{-10, -20, -30, -40, -50\}\,\mathrm{dBm}$), producing 40 distinct $(P_\mathrm{CW}, P_\mathrm{QPSK})$ combinations. 
Accounting for measured cable and propagation losses (\SI{17.5}{dB} for CW and \SI{24.2}{dB} for QPSK), the effective SIR range at the receiver spans approximately \(-33.3\,\mathrm{dB}\) to \(+46.7\,\mathrm{dB}\) (Table~\ref{tab:tx_rx_map}).

The dataset includes 8,000 mixture bursts recorded across the 40 distinct $(P_\mathrm{CW}, P_\mathrm{QPSK})$ combinations, plus 8,000 clean CW bursts recorded without interference for ground-truth reference. 
The FSQ-26 spectrum analyser has a displayed average noise level (DANL) of \SI{-145}{dBm/Hz}. 
Integrating this over the \SI{10}{MHz} baseband bandwidth and adding \SI{2}{dB} of IF ripple gives a system noise floor of approximately \SI{-102}{dBm}. 

The most challenging measurement condition occurs for a \SI{-50}{dBm} CW tone with a \SI{-10}{dBm} QPSK interferer. 
At the receiver, this corresponds to a CW level of $-50\,\mathrm{dBm} - 17.5\,\mathrm{dB} = -67.5\,\mathrm{dBm}$ and a QPSK level of $-10\,\mathrm{dBm} - 24.2\,\mathrm{dB} = -34.2\,\mathrm{dBm}$, 
resulting in a power difference of \(\sim\)33.3 dB between them. 
When combined with the requirement to resolve CW power down to the noise floor (\(-102\,\mathrm{dBm}\)), this scenario demands a dynamic range of about \( -34.2 - (-102) \approx 68 \,\mathrm{dB} \) (\(\approx 70\,\mathrm{dB}\)), which is within the analyser’s capability.

\begin{table}[H]
  \caption{Key RF parameters of the \SI{5}{GHz} capture campaign}
  \label{tab:rfparams}
  \centering
  \begin{tabular}{@{}p{4.5cm}cc@{}}
    \toprule
    \textbf{Parameter} & \textbf{Symbol} & \textbf{Value} \\
    \midrule
    System centre frequency & $f_{\mathrm c}$       & \SI{5.000}{GHz} \\
    CW tone frequency       & $f_{\mathrm{CW}}$     & \SI{5.0002}{GHz} \\
    QPSK centre frequency   & $f_{\mathrm{QPSK}}$   & \SI{5.0000}{GHz} \\
    QPSK bandwidth          & $B_{\mathrm{QPSK}}$   & \SI{4}{MHz} \\
    CW TX powers            & $P_{\mathrm{CW}}$     & \{-10, …, –50\}\,dBm \\
    QPSK TX powers          & $P_{\mathrm{QPSK}}$   & \{-10, …, –50\}\,dBm \\
    Baseband sample rate    & $f_s$                 & \SI{10}{MS/s} \\
    True SIR range (RX)     & —                     & \SIrange{-33.3}{+46.7}{dB} \\
    \bottomrule
  \end{tabular}
\end{table}

\begin{table}[H]
  \caption{Transmitted to received power mapping after calibration}
  \label{tab:tx_rx_map}
  \centering
  \begin{tabular}{lrrrrr}
    \toprule
    & \multicolumn{2}{c}{\textbf{Transmitted}} &
      \multicolumn{2}{c}{\textbf{Received}} & \textbf{SIR} \\
    Scenario & CW & QPSK & CW & QPSK & (RX) \\
    \midrule
    Weak CW   & –50 & –10 & –67.5 & –34.2 & –33.3 \\
    Strong CW & –10 & –50 & –27.5 & –74.2 & +46.7 \\
    \bottomrule
  \end{tabular}
\end{table}

\subsection{Complex-Gain Extraction}

Let \(\mathbf{x}\in\mathbb{C}^{N}\) denote a raw I/Q burst containing a CW tone, co-channel QPSK interference, and additive noise. The CW tone is modeled as a complex exponential:
\begin{equation}
  x_{\mathrm{CW}}[n] = A \cdot e^{j (2\pi f n + \phi)} = g \cdot e^{j 2\pi f n},
\end{equation}
where \(A\) is the amplitude, \(\phi\) is the phase offset, \(f\) is the normalized tone frequency, and the complex gain \(g = A e^{j\phi}\) encapsulates both amplitude and phase. To establish a ground-truth label, we estimate \(g\) from a clean CW-only burst, frequency-shifted to baseband. The gain is computed using a Hann-windowed inner product:
\begin{equation}
  g \;=\;
  \frac{\langle w,\;\mathbf{x}\,w\rangle}
       {\langle w,\;w\rangle},
  \qquad
  w[n] = \text{Hann}(N),
  \label{eq:gain_est}
\end{equation}
where \(w[n]\) is a Hann window of length \(N\), and \(\langle \cdot, \cdot \rangle\) denotes the complex inner product. The CW power in linear scale is obtained from:
\begin{equation}
  P_\mathrm{CW} = |g|^2,
\end{equation}
and in logarithmic (dBm) scale as:
\begin{equation}
  P_\mathrm{CW}^{\text{dBm}} = 10 \cdot \log_{10} \left( \frac{|g|^2}{1\ \text{mW}} \right).
\end{equation}

Only the magnitude \(|g|\) is used during training, as the phase component is unreliable due to asynchronous signal generation and acquisition. The Hann window is used to reduce spectral leakage and improve estimation accuracy. Without windowing, the finite burst length and non-integer tone frequency would cause energy from the CW to spread across frequencies, leading to biased gain estimates. The Hann window’s smooth taper confines energy to the tone’s center frequency, suppressing edge transients and improving robustness to frequency offset and additive interference.



\section{Results \& Discussion}\label{sec:results}

\begin{figure*}[!htb]
\centering
\def\imgwd{0.19\textwidth}
\def\sep{0.009\textwidth}

\captionsetup[subfigure]{font=small}

\foreach \cw in \cwpowers {%
  \noindent\makebox[\textwidth][c]{%
    \foreach \qp [count=\qi] in \qppowers {%
      \begin{subfigure}[t]{\imgwd}
        \includegraphics[width=\linewidth]{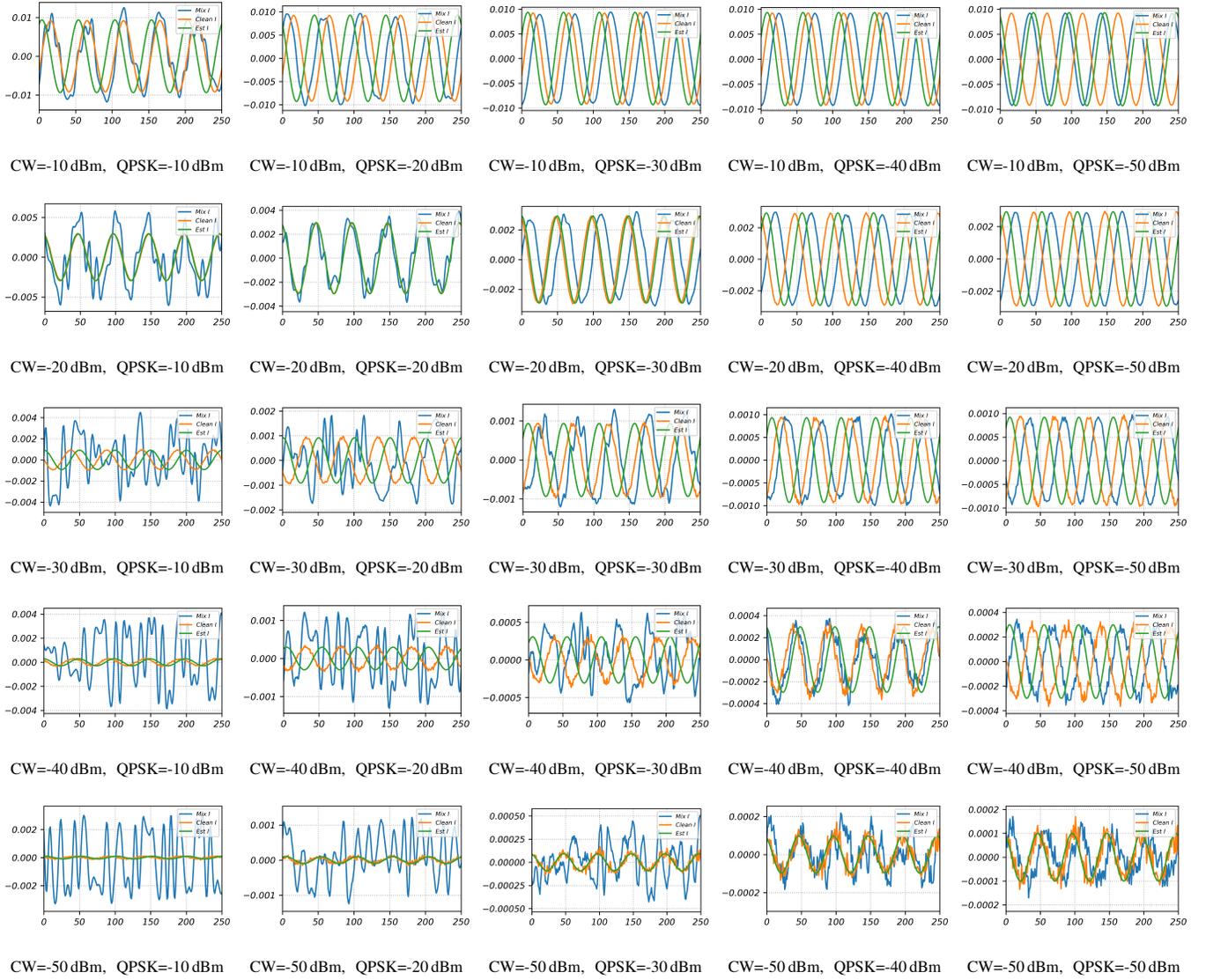}   
        {\begin{center}\scriptsize
            \caption*{\text{\scriptsize CW=\cw\,dBm,\; QPSK=\qp\,dBm}}
          
        \end{center}}
      \end{subfigure}%
      \ifnum\qi<5 \hspace{\sep}\fi
    }%
  }%
  \par\vspace{0.45\baselineskip}%
}

\caption{Time-domain I/Q bursts for every CW–QPSK power combination
(\SI{-10}{dBm} to \SI{-50}{dBm} in \SI{10}{dBm} steps).}
\label{fig:all_waves}
\end{figure*}


\begin{figure*}[!t]
\centering
\def\imgwd{0.19\textwidth}
\def\sep{0.009\textwidth}

\captionsetup[subfigure]{font=small}

\foreach \cw in \cwpowers {%
  \noindent\makebox[\textwidth][c]{%
    \foreach \qp [count=\qi] in \qppowers {%
      \begin{subfigure}[t]{\imgwd}
        \includegraphics[width=\linewidth]{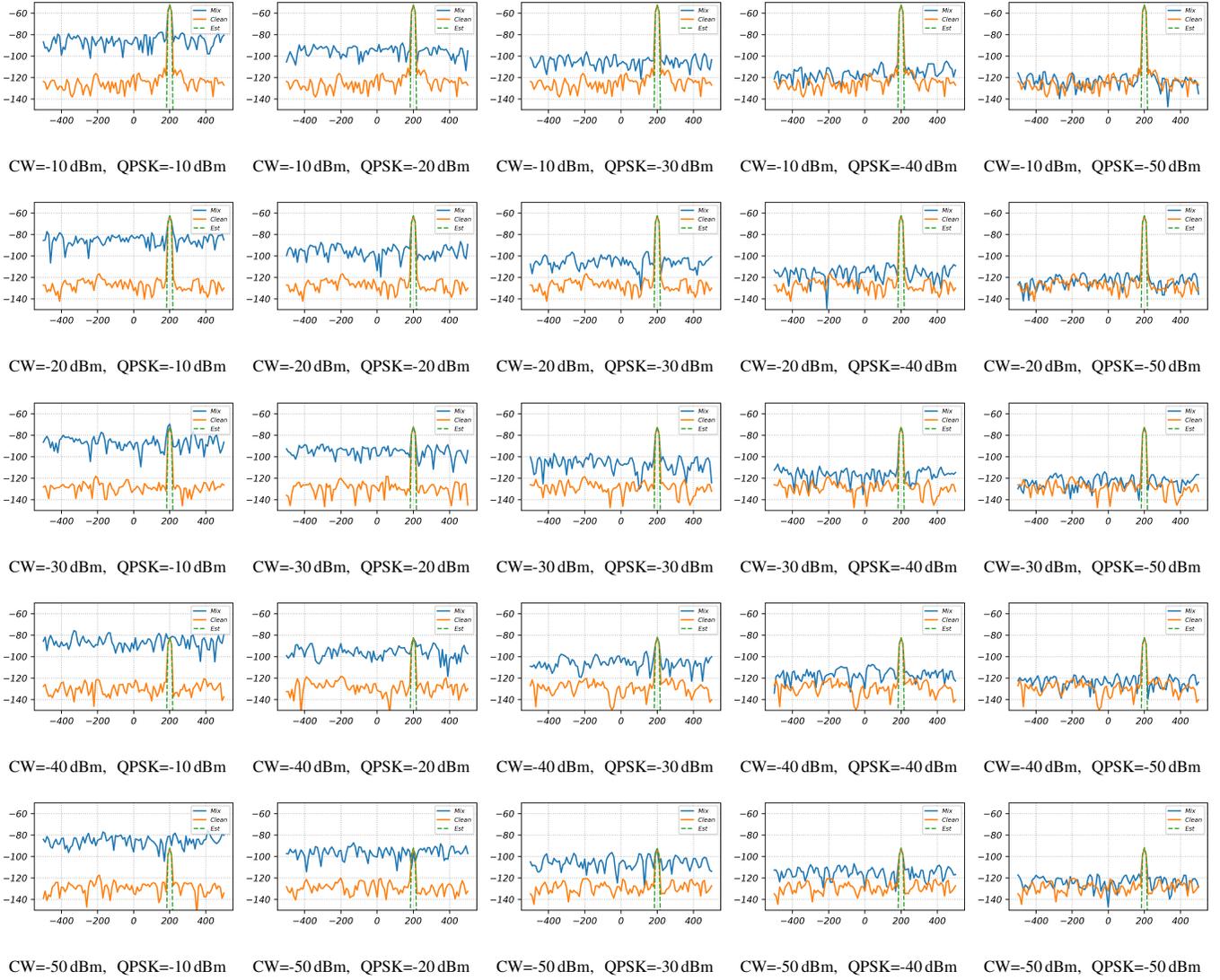}    
        {\begin{center}\scriptsize
  \caption*{\text{\scriptsize CW=\cw\,dBm,\; QPSK=\qp\,dBm}}
          \label{fig:psdcw\alph{subfigure}}
        \end{center}}
      \end{subfigure}%
      \ifnum\qi<5 \hspace{\sep}\fi
    }%
  }%
  \par\vspace{0.4\baselineskip}%
}

\caption{Power spectral density (PSD) for every CW–QPSK power combination
(\SI{-10}{dBm} to \SI{-50}{dBm} in \SI{10}{dBm} steps).}
\label{fig:all_psds}
\end{figure*}

\vspace{0.6\baselineskip}

\subsection{Time-Domain and Spectral Characteristics}

Figs.~\ref{fig:all_waves} and \ref{fig:all_psds} illustrate the time-domain I/Q bursts and corresponding power spectral densities (PSDs) for every CW–QPSK power combination in the test set. In each plot, the \emph{model-estimated CW component} is overlaid with the measured signal, allowing direct visual inspection of how well the estimator recovers the carrier under different interference levels.

In the time-domain plots (Fig.~\ref{fig:all_waves}), high-SIR conditions (top-left panels) show the estimated CW waveform closely matching the measured signal, with minimal residual error. As the QPSK interference power increases, visible amplitude and phase perturbations appear, yet the estimator tracks the underlying sinusoid with small deviations until very low SIR. At the most challenging cases (bottom-right), the true CW is almost entirely buried under the interferer, and the model output shows small but consistent bias.

The PSD plots in Fig.~\ref{fig:all_psds} offer the frequency-domain perspective. Here, the estimated CW power is plotted alongside the measured PSD, highlighting how the model identifies the carrier peak even when it is partially or fully masked by the QPSK spectrum. For high SIR, the CW peak dominates and aligns perfectly between the estimate and the measurement. At low SIR, the QPSK side-lobes merge with the carrier bin, making FFT-based isolation unreliable—yet the neural estimators still identify the correct carrier location and approximate its power.

Together, these figures illustrate not only the range of interference conditions encountered but also the estimator’s resilience, directly connecting the raw signal characteristics to the quantitative accuracy results presented next.


\begin{figure*}[!htb]
  \begin{subfigure}[t]{0.48\textwidth}
    \includegraphics[width=\linewidth]{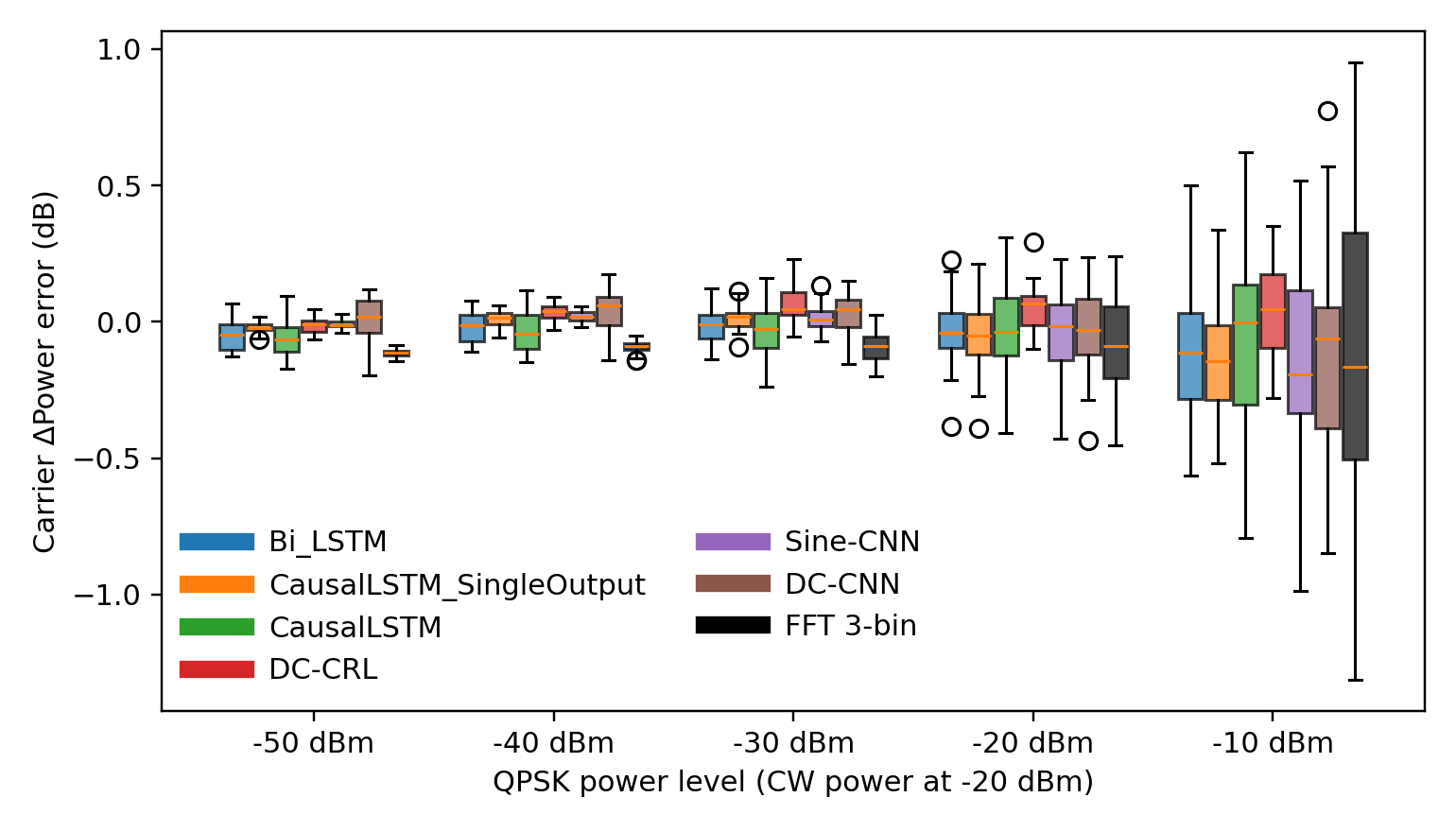}
    \caption{CW = \SI{-20}{dBm}}
    \label{fig:err_vs_QPSK_20}
  \end{subfigure}\hfill
  \begin{subfigure}[t]{0.48\textwidth}
    \includegraphics[width=1.05\linewidth]{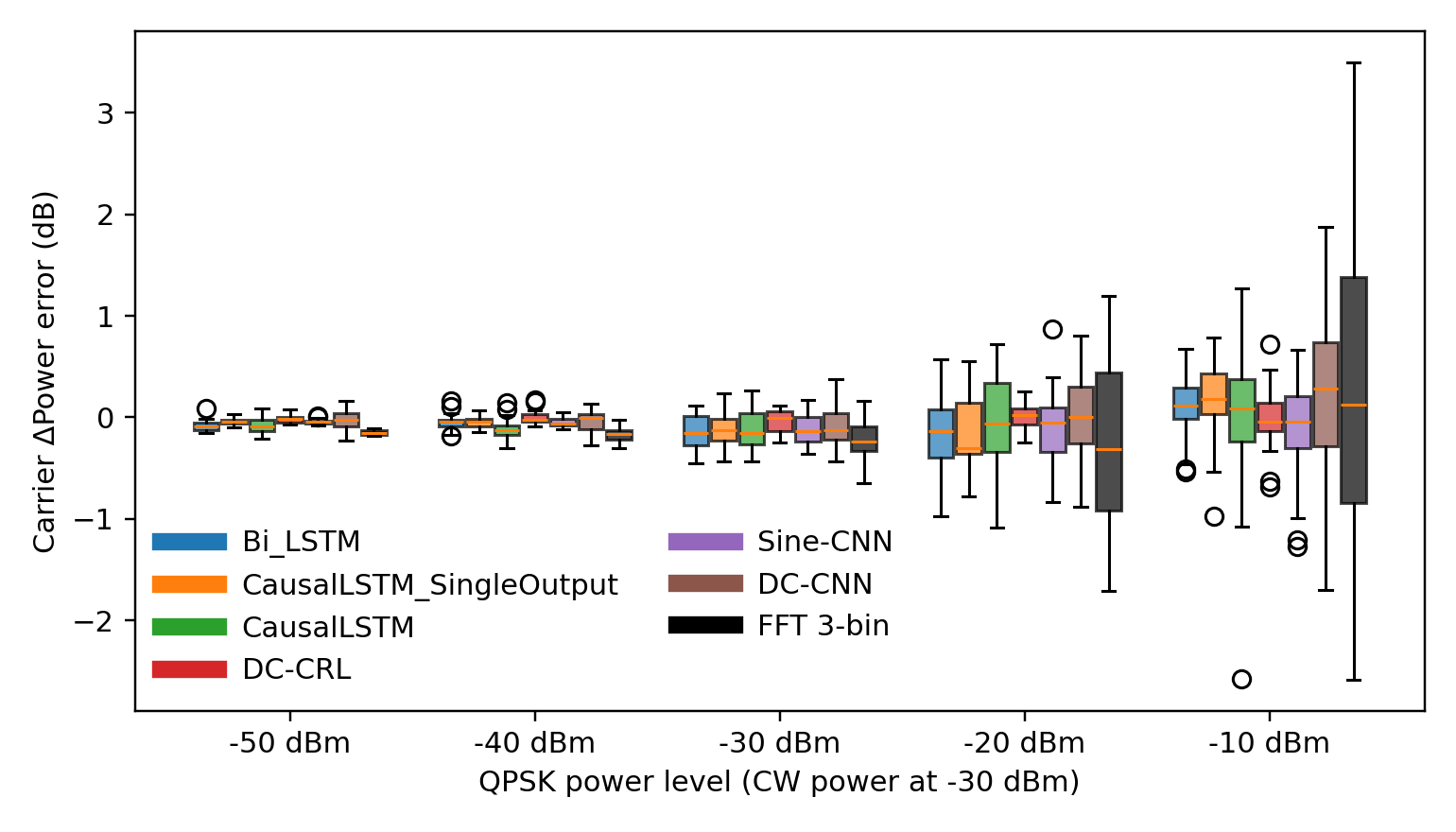}
    \caption{CW = \SI{-30}{dBm}}
    \label{fig:err_vs_QPSK_30}
  \end{subfigure}

  \begin{subfigure}[t]{0.48\textwidth}
    \includegraphics[width=\linewidth]{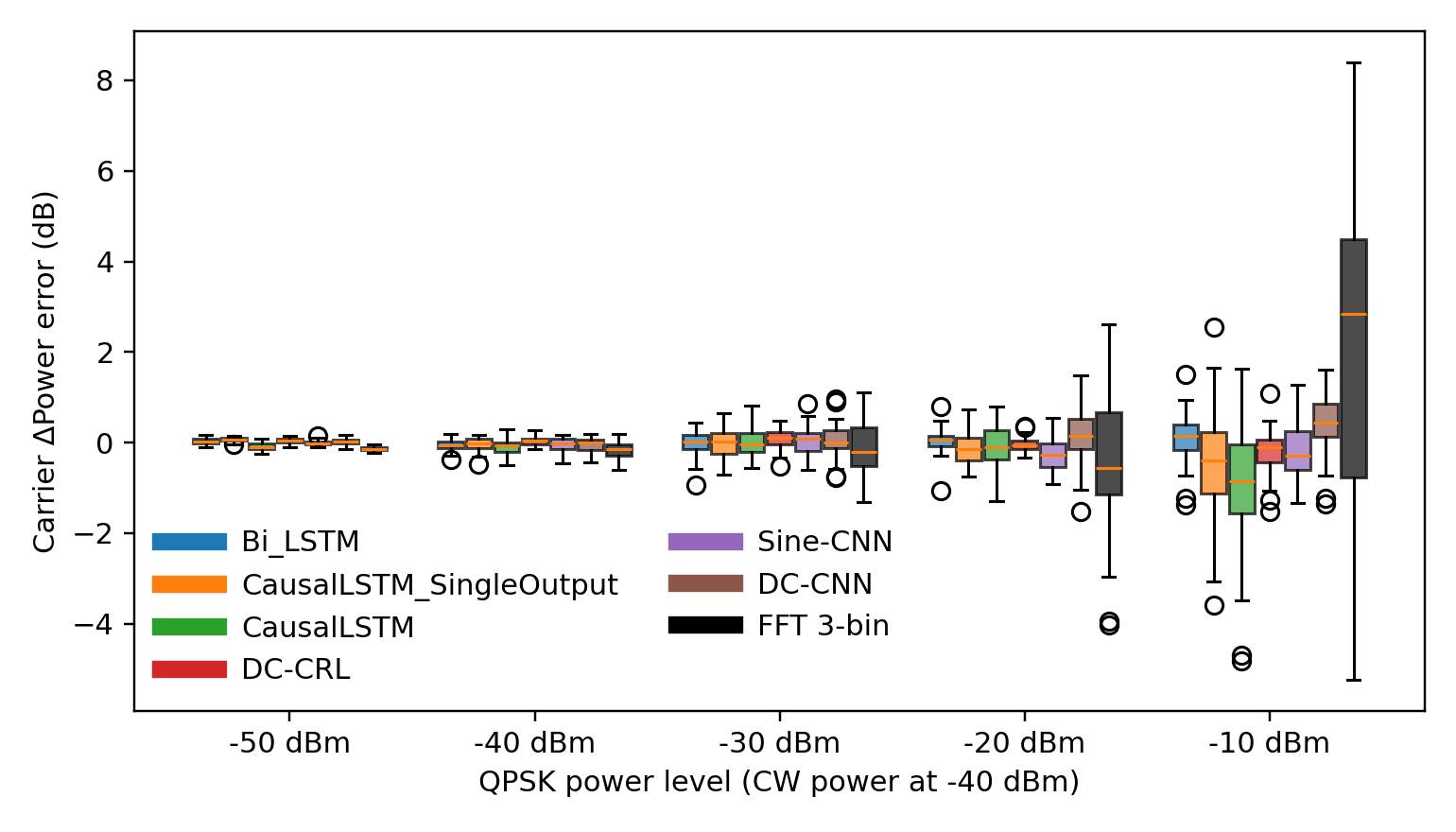}
    \caption{CW = \SI{-40}{dBm}}
    \label{fig:err_vs_QPSK_40}
  \end{subfigure}\hfill
  \begin{subfigure}[t]{0.48\textwidth}
    \includegraphics[width=\linewidth]{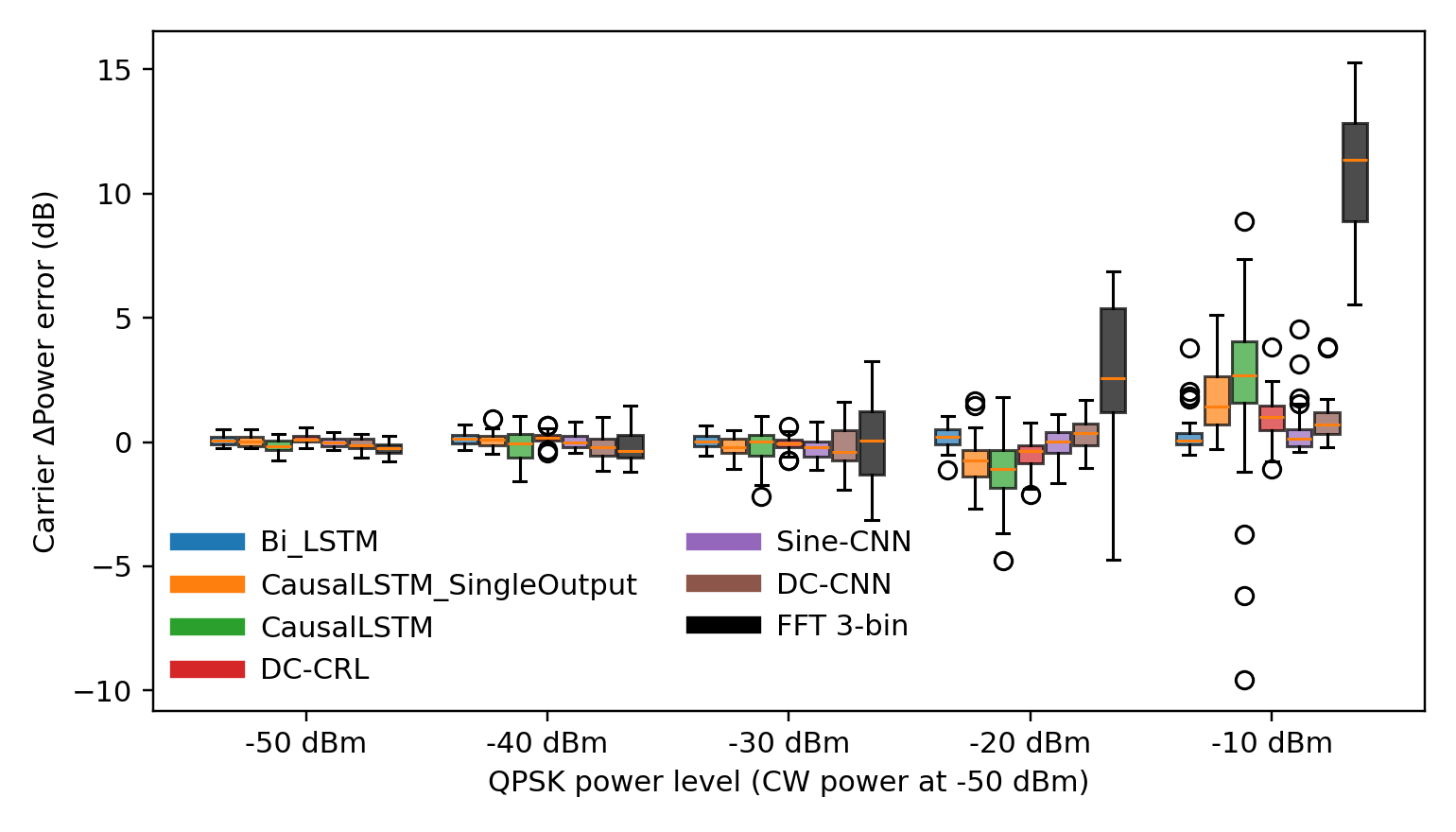}
    \caption{CW = \SI{-50}{dBm}}
    \label{fig:err_vs_QPSK_50}
  \end{subfigure}

  \caption{Carrier–power error ($\Delta P_{\text{CW}}$) of all estimators versus QPSK interference level for four CW power levels.  The boxplots summarise 30 test bursts per QPSK power level: the solid line marks the median, boxes span the inter-quartile range (IQR), and whiskers extend to $1.5\times$ IQR.}
  \label{fig:err_vs_QPSK}
\end{figure*}

\subsection{Carrier–Power Error versus Interference Level}
Fig.~\ref{fig:err_vs_QPSK} presents boxplots of the carrier–power error $\Delta P_{\mathrm{CW}}$ across QPSK power levels for four representative CW power settings. For high CW powers (Figs.~\ref{fig:err_vs_QPSK_20} and~\ref{fig:err_vs_QPSK_30}), all neural estimators maintain median error within $\pm0.2$ dB for SIR down to approximately $-10$ dB, with the FFT baseline showing larger dispersion as interference increases. For weaker CW tones (Figs.~\ref{fig:err_vs_QPSK_40} and~\ref{fig:err_vs_QPSK_50}), the interquartile range broadens more noticeably for all methods under strong interference, but the DC-CRL and Bi-LSTM remain the most consistent performers.

These results demonstrate that while recurrent-augmented convolutional models deliver slightly improved robustness at very low SIR, the proposed lightweight DC-CNN still tracks the CW amplitude within $\approx$1 dB in most tested conditions, making it well-suited for embedded deployment where computational resources are constrained.

\subsection{Carrier Power Estimation Accuracy Across the Full SIR Range}

\begin{figure}[t]
  \centering
  \includegraphics[width=\linewidth]{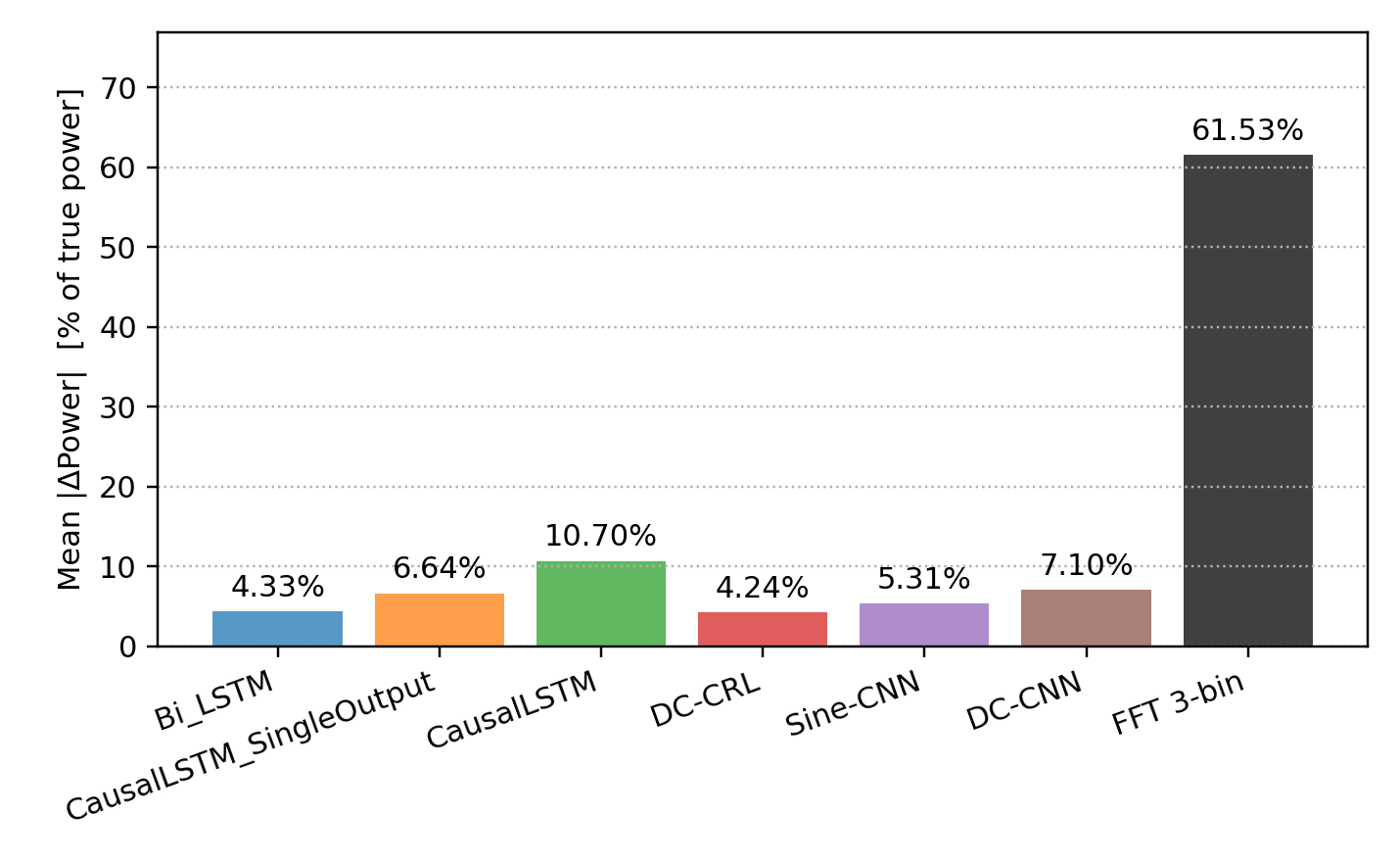}
  \caption{Mean carrier-power estimation error (MAE$_{\%}$) across the full SIR range ($-40$\,dB to $+40$\,dB). All learned models stay below 8\% average error.}
  \label{fig:5b}
\end{figure}

\begin{figure}[t]
  \centering
  \includegraphics[width=\linewidth]{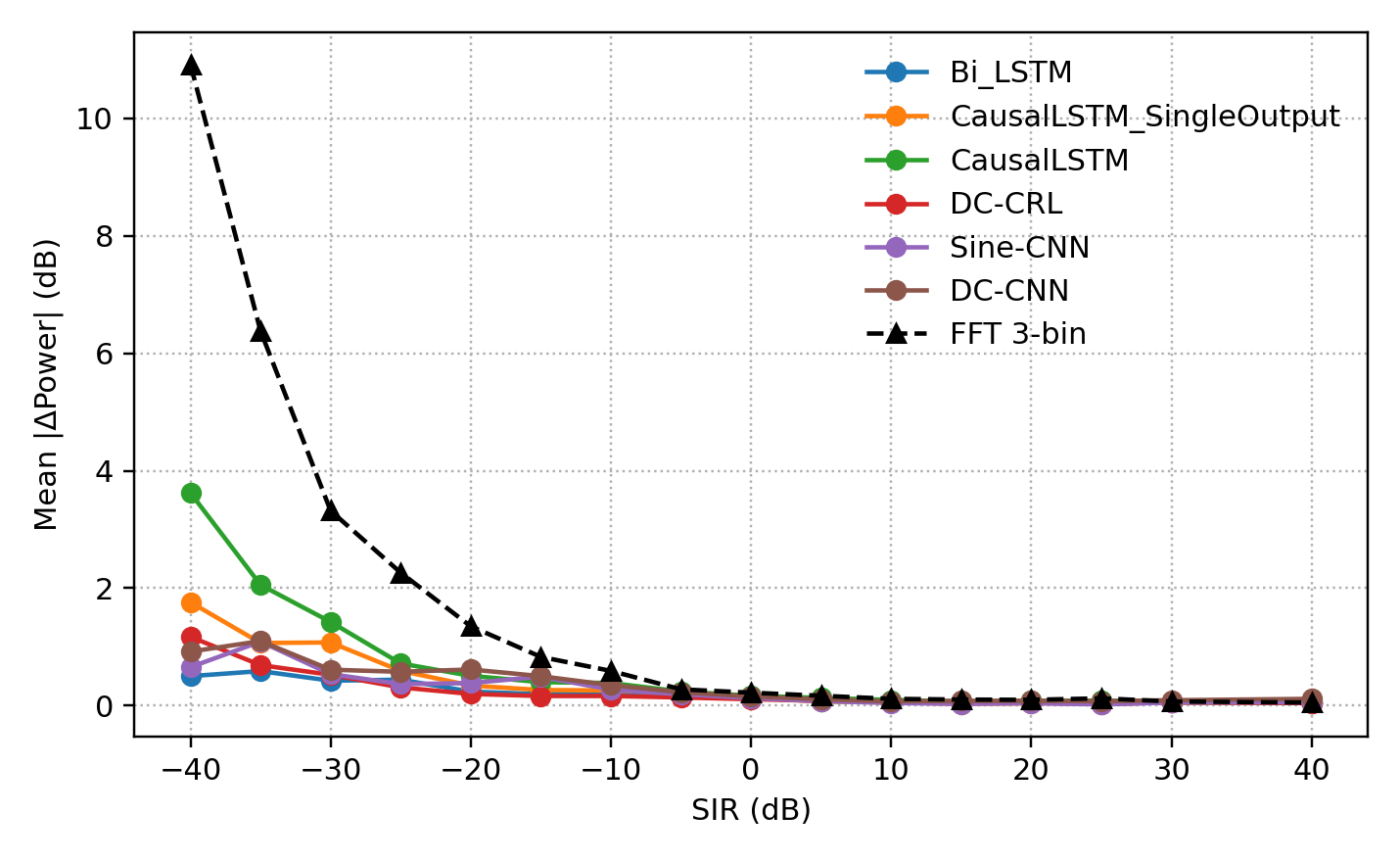}
  \caption{MAE (dB) versus nominal SIR. Shaded bands denote $\pm1\sigma$
           across bursts. All models except one remain under 2\,dB error across the full SIR range.}
  \label{fig:mae_vs_sir}
\end{figure}

Figure~\ref{fig:5b} compares the mean carrier--power estimation error (MAE$_{\%}$) of all estimators over the complete SIR sweep from $-40\,\mathrm{dB}$ to $+40\,\mathrm{dB}$. All learned models maintain average errors below $8\%$, with the best performance achieved by the DC-CRL hybrid ($4.24\%$), closely followed by the Bi-LSTM ($4.33\%$) and Sine-CNN ($5.31\%$). The proposed DC-CNN records a modestly higher error of $7.10\%$ while preserving a lightweight architecture suitable for embedded inference. In contrast, the classical 3-bin FFT baseline exhibits a substantial error of $61.53\%$, underscoring its vulnerability to QPSK interference.

Figure~\ref{fig:mae_vs_sir} presents the mean absolute error in dB versus nominal SIR. Across most of the SIR range, all neural models---except CausalLSTM---stay well below the $2\,\mathrm{dB}$ error threshold. The CausalLSTM’s higher error at low SIR stems from its sequence-to-sequence prediction burden, which requires reconstructing the full 1{,}000-sample I/Q burst rather than directly regressing the complex gain. The FFT baseline’s error grows rapidly for $\mathrm{SIR} \le -20\,\mathrm{dB}$, exceeding $10\,\mathrm{dB}$ under the strongest interference conditions. In contrast, the top-performing models exhibit smooth degradation and maintain median errors within $0.2\,\mathrm{dB}$ for $\mathrm{SIR} \ge -10\,\mathrm{dB}$.

Overall, these results confirm that the proposed DC-CNN offers a favorable trade-off between estimation accuracy and model complexity, delivering sub-$1\,\mathrm{dB}$ accuracy across most operational SIR conditions.


\section{Conclusion \& Future Work}\label{sec:conclusion}

This work presents DC-CNN, an interference-robust CW power estimator tailored for in-situ, UAV-enabled, large-aperture antenna radiation characterization. Operating reliably over a SIR range of $-33.3\,\mathrm{dB} \le \mathrm{SIR} \le +46.7\,\mathrm{dB}$, the proposed model achieves sub-\SI{1}{dB} mean absolute error (MAE) in CW power estimation under strong in-band QPSK interference. Its lightweight design enables deployment on embedded UAV platforms, facilitating rapid, wide-area radiation measurements without the need for bulky ground-based equipment.

Future work will explore multi-tone regression to allow simultaneous multi-frequency measurements, thereby increasing measurement speed for frequency band coverage and reducing the required UAV flight time. Additional directions include integrating few-shot, on-device self-supervised adaptation to mitigate time-varying oscillator drift and thermal effects, as well as extending interference robustness to modulation schemes beyond QPSK. Finally, the designed algorithms will be developed on embedded UAV hardware, advancing toward fully autonomous in-situ antenna diagnostics.

\section*{Acknowledgment}

The authors thank to Kim Olesen for providing access to laboratory equipment.

\end{document}